\documentclass[12pt]{article}

\usepackage{color}
\definecolor{darkblue}{rgb}{0.1,0.1,.7}
\usepackage[colorlinks, linkcolor=darkblue, citecolor=darkblue, urlcolor=darkblue, linktocpage]{hyperref}
\usepackage[]{amsmath}
\usepackage[]{graphicx}
\usepackage[]{latexsym}
\usepackage{slashed,graphicx,color,amsmath,amssymb}
\usepackage[mathscr]{eucal}
\usepackage{mathrsfs}
\usepackage{geometry}
\usepackage[margin=10pt,font=small,labelfont=bf]{caption}
\usepackage{amscd}
\usepackage{bm}
\usepackage{xcolor}
\usepackage{upgreek}
\usepackage[square, comma, sort&compress,numbers]{natbib}
\usepackage[all,cmtip]{xy}
\numberwithin{equation}{section}
\newcommand{\tr}{\mathrm{Tr}\,}

\newcommand{\rH}{\mathrm{H}}

\newcommand{\rT}{\mathrm{T}}

\begin{document}
\vspace*{-.6in} \thispagestyle{empty}
\vspace{.4in} {\Large
\begin{center}
{\bf Diagonal Form Factors and Hexagon Form Factors II\\Non-BPS Light Operator}
\end{center}}
\vspace{.2in}
\begin{center}
Yunfeng Jiang
\\
\vspace{.3in}
\small{\textit{Institut f{\"u}r Theoretische Physik,
ETH Z{\"u}rich}},\\
\small{\textit{
Wolfgang Pauli Strasse 27,
CH-8093 Z{\"u}rich, Switzerland}
}

\vspace{.1in}
\small{\texttt{jinagyf2008@gmail.com}}
\end{center}

\vspace{.3in}

\begin{abstract}
\normalsize{We study the asymptotic volume dependence of the heavy-heavy-light three-point functions in the $\mathcal{N}=4$ Super-Yang-Mills theory using the hexagon bootstrap approach, where the volume is the length of the heavy operator. We extend the analysis of our previous short letter \cite{Jiang:2015bvm} to the general case where the heavy operators can be in any rank one sector and the light operator being a generic non-BPS operator. We prove the conjecture of Bajnok, Janik and Wereszczynski \cite{Bajnok:2014sza} up to leading finite size corrections.}
\end{abstract}

\vskip 1cm \hspace{0.7cm}

\newpage

\setcounter{page}{1}
\begingroup
\hypersetup{linkcolor=black}
\setcounter{tocdepth}{1}
\tableofcontents
\endgroup

\section{Introduction}
\label{sec:intro}
Form factor bootstrap program \cite{Smirnov:1992vz} is a powerful method to obtain non-perturbative results of correlation functions in integrable systems. The $\mathcal{N}=4$ Super-Yang-Mills theory in the large $N_c$ limit is a new kind of integrable system with very rich structure. In recent years, there has been many solid progress in the computation of three-point functions in the planar $\mathcal{N}=4$ SYM theory, both at weak and strong coupling \cite{old,tailoring,freezing,oneloop,SoV,othersector,strongC,twist2,short,OPEPotugal}. In order to tackle this problem at finite coupling, it is desirable to relate the three-point functions to form factors and apply the bootstrap methodology.\par

There are at least three proposals in this direction so far. Klose and McLaughlin proposed a set of bootstrap axioms for the worldsheet form factors \cite{Klose:2012ju}. This is a direct generalization of the form factor bootstrap program in 2d integrable field theories. However, due to the non-relativistic nature of the light-cone gauge fixed string theory and the complicated spectrum of the theory, it is highly challenging to solve the bootstrap axioms. Also, relating the worldsheet form factors to three-point functions is a non-trivial problem.\par

Inspired by the structure of lightcone string field theory, which has been used to calculate three-point functions in the BMN regime, Bajnok and Janik proposed a set of axioms for the so-called generalized Neumann coefficient \cite{Bajnok:2015hla}. This object can be defined for any integrable field theories and is obtained by taking a special decompactification limit of the structure constant. In contrary to the usual form factors in integrable field theories, the generalized Neumann coefficient corresponds to form factors of \textit{non-local} operators. This fact modifies the form factor bootstrap axioms by some extra phase factors. Again the set of axioms is quite challenging to solve, but some progress has been made recently in \cite{Bajnok:2015ftj}. At weak coupling, similar ideas have led to the proposal of the spin vertex formalism \cite{Jiang:2014cya,Jiang:2014gha,Kazama:2014sxa}.\par

Very recently, Basso, Komatsu and Vieira \cite{Basso:2015zoa} proposed a different method called the hexagon bootstrap program. In this method, one cut the three-point function, which is represented by a pair of pants, into two more fundamental objects called the hexagons or the hexagon form factors. The authors of \cite{Basso:2015zoa} proposed a set of bootstrap axioms for the hexagon form factor which can be solved explicitly. Gluing back the two hexagons by taking into account the mirror excitations, one obtains the structure constant. This method has been verified by many non-trivial checks \cite{Basso:2015zoa,Eden:2015ija,Basso:2015eqa}.\par

Apart form these proposals, there is yet another way of relating form factors to a special type of three-point functions called the heavy-heavy-light (HHL) three-point function. Here heavy (light) means the quantum number of the operator is large (small). This type of three-point function is first investigated in the dual string theory in \cite{Zarembo:2010ab,Costa:3pt}. It can be seen as a kind of ``perturbation'' of classical string solutions with light supergravity modes. If we regard the two heavy operators as incoming and outgoing states and the light operator as some operator sandwiched between these states, then the HHL three-point function can be seen as a diagonal form factor or the mean value of the light operator in the state corresponding to the heavy operator. This idea is made more concrete in \cite{Bajnok:2014sza}.\par

The form factor bootstrap method gives us non-perturbative result in infinite volume. In the context of three-point function, the ``volume'' is the length of the operator and should be finite. It is therefore an important question to take into account the volume corrections. There are in general two type of volume corrections. The first type is called \textit{asymptotic} volume correction, which takes the form of polynomials of $1/L$. It originates from imposing the periodic boundary condition which changes the quantization condition of the excitations. The second type is called \textit{wrapping corrections} or finite volume corrections, which is due to the propagation of virtual particles and takes an exponential form $e^{-E\,L}$ where $E$ is the energy of the virtue particle and $L$ is the length that it propagates. While the asymptotic volume corrections can be taken into account in a systematic manner, it is notoriously hard to take into account the wrapping corrections.\par

Based on previous studies in the 2d integrable field theories \cite{Pozsgay:2007gx}, Bajnok, Janik and Wereszczyski proposed a conjecture concerning the asymptotic volume dependence of the HHL structure constant at any coupling. This conjecture was checked at strong coupling by the same authors for several examples and at weak coupling in \cite{Hollo:2015cda} in the $\mathfrak{su}(2)$ sector. Using the hexagon form factor approach, the BJW conjecture is also checked at finite coupling in the $\mathfrak{su}(2)$ sector for the light operator being the BMN vacuum \cite{Jiang:2015bvm}. In this paper, we generalize the result of \cite{Jiang:2015bvm} and show that the BJW conjecture is valid for non-BPS light operator. We prove the conjecture for all the rank one sectors, namely $\mathfrak{su}(2)$, $\mathfrak{sl}(2)$ and $\mathfrak{su}(1|1)$ sectors. As in \cite{Jiang:2015bvm}, due to the fact that it is not yet clear how to take into account all the mirror excitations, we restrict our proof to only the physical excitations.\par

The rest of the paper is organized as follows. In section \ref{sec:setup}, we describe the set-up of the problem. In section \ref{sec:hexagon}, we present a method to check the validity of BJW conjecture directly for few excitations. When the rapidities of two excitations on the two physical edges coincide, they will decouple and the hexagon form factor is proportional to the one without these two excitations. In section \ref{sec:factorization}, we study this decoupling limit in detail. We call the relation of the hexagons before and after decoupling the factorization properties. In section \ref{sec:proof}, we prove the BJW conjecture up to mirror excitations. In section \ref{sec:inf}, we give some comments on the infinite volume form factors which appear in the BJW conjecture. We conclude in section \ref{sec:conc} and discuss future directions to explore. Some complementary details are presented in the appendices.

\section{The set-up}
\label{sec:setup}
\begin{figure}[h!]
\begin{center}
\includegraphics[scale=0.5]{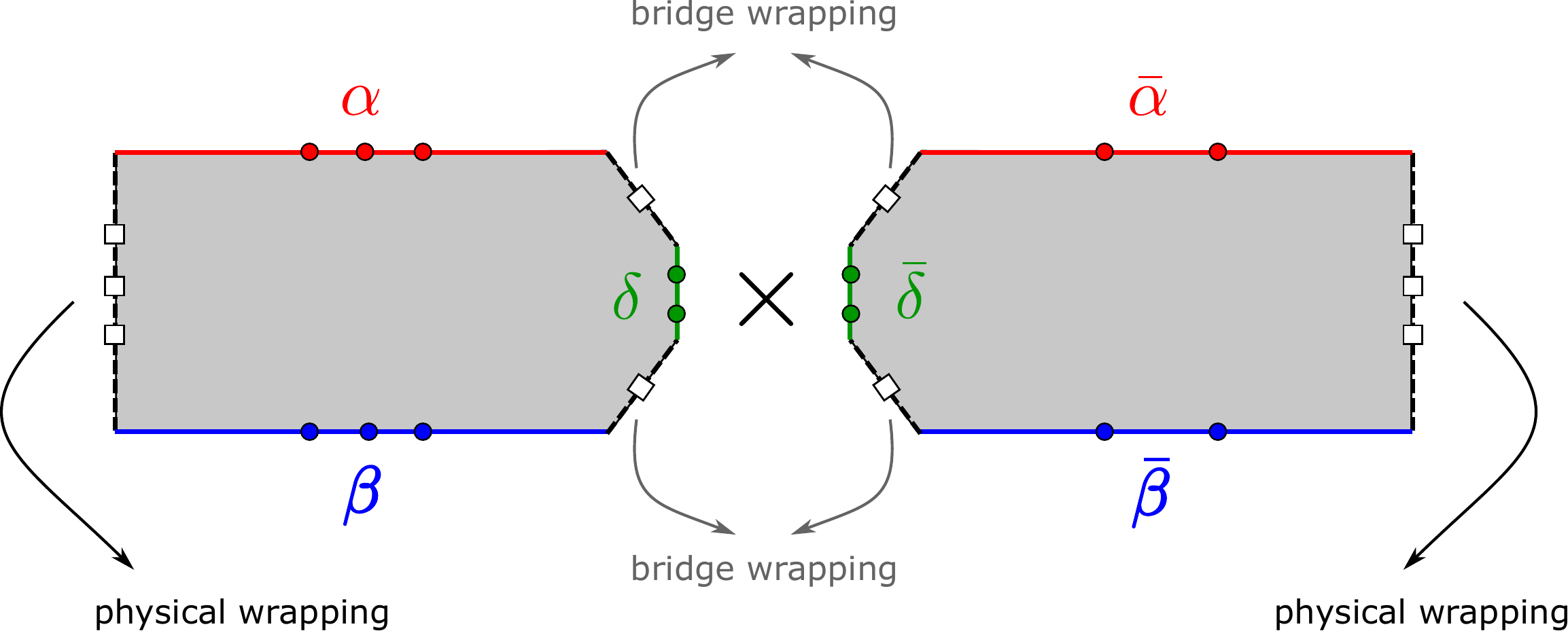}
\caption{The arrangement of excitations on the two hexagons.}
\label{fig:CH}
\end{center}
\end{figure}

In this section, we give the set-up of our problem. For HHL three-point functions, the two heavy operators $\mathcal{O}_1$ and $\mathcal{O}_2$ are conjugate to each other. They can be chosen in any of the three rank one sectors, namely the $\mathfrak{su}(2)$, $\mathfrak{sl}(2)$ and $\mathfrak{su}(1|1)$ sectors. The excitations in these three sectors are scalars, (covariant) derivatives and fermions, respectively. We denote a generic excitation by $\chi$ and its conjugate by $\bar{\chi}$. There are 8 pairs of excitations:
\begin{center}
\begin{tabular}{|c|c|c|c|c|c|c|c|c|}
  \hline
  $\chi$       & $\Phi_{1\dot{1}}$            & $\Phi_{1\dot{2}}$            & $\mathcal{D}_{3\dot{3}}$            &$\mathcal{D}_{3\dot{4}}$  & $\Psi_{1\dot{3}}$ & $\Psi_{2\dot{3}}$ & $\Psi_{1\dot{4}}$ & $\Psi_{2\dot{4}}$ \\
  \hline
  $\bar{\chi}$ & ${\Phi}_{2\dot{2}}$ & ${\Phi}_{2\dot{1}}$ & ${\mathcal{D}}_{4\dot{4}}$ &${\mathcal{D}}_{4\dot{3}}$ & ${\Psi}_{2\dot{4}}$ & ${\Psi}_{1\dot{4}}$ & ${\Psi}_{2\dot{3}}$ & ${\Psi}_{1\dot{3}}$ \\
  \hline
\end{tabular}
\end{center}
The polarizations of the excitations of the two heavy operators are chosen such that $\mathcal{O}_1:\,\chi$ and $\mathcal{O}_2:\,\bar{\chi}^{2\gamma}$ so that by performing $2\gamma$ transformations of the excitations on $\mathcal{O}_2$ to the edge of $\mathcal{O}_1$ the two sets of excitations are conjugated to each other.
Let us denote the length and the number of excitations of the heavy operators by $L$ and $N$. The two heavy operators take the following form
\begin{align}
\mathcal{O}_1=\tr Z^{L-N}\chi^N+\cdots,\qquad \mathcal{O}_2=\tr\bar{Z}^{L-N}\bar{\chi}^N+\cdots.
\end{align}
We denote the two sets of rapidities of $\chi$ and $\bar{\chi}$ by $\mathbf{u}=\{u_1,\cdots,u_N\}$ and $\mathbf{v}=\{v_1,\cdots,v_N\}$ respectively. The length of the third operator is denoted by $2l_0$, where $l_0\ll L$. In the previous paper \cite{Jiang:2015bvm}, the light operator is taken to be the BPS operator $\tr\tilde{Z}^{2l_0}$ with $\tilde{Z}=Z+\bar{Z}+Y-\bar{Y}$. In the current paper, we will consider general non-BPS operators and put excitations on the BMN vacuum. The set of excitations of the light operator is denoted by\footnote{If there are more than one type of excitations, we need to use the nested Bethe ansatz and take proper linear combinations of the excitations in order the third operator to have well-defined scaling dimension. In that case, what we consider is one of the terms in the linear combination. We show that the BJW conjecture holds for each term and thus holds for the whole operator.}
\begin{align}
\mathcal{X}_{\mathbf{A}\dot{\mathbf{A}}}(\mathbf{w})=\{\mathcal{X}_{A_1\dot{A}_1}(w_1),\mathcal{X}_{A_2\dot{A}_2}(w_2),\cdots,\mathcal{X}_{A_n\dot{A}_n}(w_n)\}.
\end{align}
where $\mathcal{X}_{A_k\dot{A}_k}$ denotes a generic excitation. According to the hexagon approach \cite{Basso:2015zoa}, the asymptotic structure constant with three non-BPS operators is given by the following sum-over-partition formula
\begin{align}
\label{eq:CSoP}
C_{123}^{\bullet\bullet\bullet}=\sum_{\substack{\alpha\cup\bar{\alpha}=\mathbf{u}\\ \beta\cup\bar{\beta}=\mathbf{v}\\ \delta\cup\bar{\delta}=\mathbf{w}}}
(-1)^{|\bar{\alpha}|+|\bar{\beta}|+|\bar{\delta}|}\omega_{l_{31}}(\alpha,\bar{\alpha})\omega_{l_{12}}(\beta,\bar{\beta})\omega_{l_{23}}(\delta,\bar{\delta})
\times \rH(\alpha|\delta|\beta)\,\rH(\bar{\beta}|\bar{\delta}|\bar{\alpha})
\end{align}
The arrangement of excitations is depicted in figure\,\ref{fig:CH}.
\\
For our set-up, we have $L_1=L_2=L$, $L_3=2l_0$ and
\begin{align}
l_{12}=L-l_0,\quad l_{23}=l_0,\quad l_{31}=l_0.
\end{align}
For later convenience, we denote $l=L-l_0$. The explicit form of the splitting factor depends on the ordering of the excitations. However, the normalized structure constant (\ref{eq:C123}) does not depend on the ordering. We choose the same ordering as in \cite{Jiang:2015bvm}, namely the rapidities $\mathbf{u}$ are reverse ordered. We use the Bethe Ansatz Equations (BAE) for the first splitting factor $\omega_{l_{31}}(\alpha,\bar{\alpha})$ and write
\begin{align}
\omega_{-l}(\alpha,\bar{\alpha})=\prod_{u_j\in\bar{\alpha}}(e^{-ilp(u_j)}\prod_{\substack{u_i\in\alpha\\i>j}}S(u_i,u_j))
\end{align}
The second splitting factor $\omega_{l_{12}}(\beta,\bar{\beta})$ can be written as
\begin{align}
\omega_{l}(\beta,\bar{\beta})=\prod_{v_j\in\bar{\beta}}(e^{ilp(v_j)}\prod_{\substack{v_i\in\beta\\i>j}}S(v_j,v_i))
\end{align}
We do not need the explicit form of the third splitting factor\footnote{If there are more than one type of excitations for the light operator, the splitting factor can be a matrix instead of a scalar function.} and we simply denoted by $\omega_{l_0}(\delta,\bar{\delta})$.
The un-normalized HHL structure constant is defined as the diagonal limit of $C_{123}^{\bullet\bullet\bullet}$
\begin{align}
\mathcal{C}_{\text{HHL}}=\lim_{\mathbf{v}\to\mathbf{u}}C_{123}^{\bullet\bullet\bullet}
\end{align}
The quantity we want to study is the following \textit{normalized} HHL structure constant
\begin{align}
\label{eq:C123}
C_{\text{HHL}}=\frac{1}{\prod_{i=1}^N\,a_\chi(u_i)}\frac{1}{\rho_{\chi,N}(\mathbf{u})}\,\mathcal{C}_{\text{HHL}}.
\end{align}
The normalization constant $a_\chi(u)$ is defined
\begin{align}
a_\chi(u)=(-1)^{\dot{f}_{\bar{\chi}^{2\gamma}}f_\chi+1}n_\chi\,\mu(u)
\end{align}
where $f$ and $\dot{f}$ is the fermionic number of the un-dotted and dotted indices of the excitations, $\mu(u)$ is the measure introduced in \cite{Basso:2015zoa} and $n_\chi$ is a simple number which will be defined in section \ref{sec:factorization}. In (\ref{eq:C123}), $\rho_{\chi,N}(\mathbf{u})$ is the asymptotic Gaudin determinant of the $\mathfrak{su}(2)$, $\mathfrak{sl}(2)$ and $\mathfrak{su}(1|1)$ sector for $\chi$ being scalars, derivatives and fermions, respectively. The asymptotic Gaudin determinant is proportional to the norm of the Bethe state and is given by
\begin{align}
\rho_{\chi,N}(\mathbf{u})=\det_{j,k}\frac{\partial}{\partial u_j}\Phi_{\chi,k},\qquad \Phi_k=p(u_k)L-i\sum_{l\ne k}\log S_{\chi}(u_k,u_l).
\end{align}
where $S_{\chi}(u,v)$ is the $S$-matrix in the corresponding subsectors. When we consider a subset $\alpha\subset\mathbf{u}$, we can define two quantities related to $\rho_N(\mathbf{u})$. We define $\rho^s_{|\alpha|}(\alpha)$ as the Gaudin determinant with respect to the rapidities $u_j\in\alpha$ and $\rho^c_N(\alpha)$ as the diagonal minor of the Gaudin determinant $\rho_N(\mathbf{u})$ with respect to $u_j\in\alpha$. While $\rho^s_{|\alpha|}(\alpha)$ depends only on the rapidities in the set $\alpha$, $\rho^c_N(\alpha)$ depends on all the rapidities $\mathbf{u}$. We will prove that
\begin{align}
\label{eq:PT}
C_{\text{HHL}}=\frac{1}{\rho_{\chi,N}(\mathbf{u})}\sum_{\alpha\cup\bar{\alpha}=\mathbf{u}}{F}^{\mathbf{w},s}_{|\alpha|}({\alpha})\rho^s_{\chi,|\bar{\alpha}|}(\bar{\alpha})
\end{align}
where ${F}^{\mathbf{w},s}_{|\alpha|}(\alpha)$ is some well-defined quantity in infinite volume, which we shall call the infinite volume form factor. A theorem in \cite{Pozsgay:2007gx} states that (\ref{eq:PT}) has another equivalent expansion in terms of $\rho^c_N(\alpha)$
\begin{align}
\label{eq:PTc}
C_{\text{HHL}}=\frac{1}{\rho_{\chi,N}(\mathbf{u})}\sum_{\alpha\cup\bar{\alpha}=\mathbf{u}}{F}^{\mathbf{w},c}_{|\alpha|}({\alpha})\rho^c_{\chi,N}(\bar{\alpha})
\end{align}
where ${F}^{\mathbf{w},c}_{|\alpha|}(\alpha)$ is different from ${F}^{\mathbf{w},s}_{|\alpha|}(\alpha)$ in general, but they are related by the relations given in \cite{Pozsgay:2007gx}. The fact that we have two expansions reveals the ambiguity of the diagonal form factor in the infinite volume. Nevertheless, the finite volume form factor $C_{\text{HHL}}$ is unambiguously defined.\par

Finally we comment on the mirror excitations. In order to obtain the complete result of the structure constant, we need to take into account all the mirror excitations on the three mirror edges, as is shown in figure\,\ref{fig:CH}. The mirror excitations on the opposite edge to the edge of the light operator corresponds to the physical wrapping corrections, which are of order $e^{-E\,L}$ and can be neglected safely since we are considering the large $L$ limit. The mirror excitations on the edges that are adjacent to the edge of the light operator leads to the so-called bridge wrapping corrections, which is of order $e^{-E\,l_0}$. Since $l_0$ is finite, we should take into account all the mirror excitations on the adjacent mirror edges. However, in the hexagon approach, when two mirror excitations on the two adjacent edges coincide, there is a double pole in the integrand and so far it is not yet clear how to deal with this divergence. Due to this restriction, we will not consider any mirror excitations in this paper and leave this question for future investigations. We stress here that our proof in this paper is only up to mirror excitations.

\section{A direct check of BJW conjecture}
\label{sec:hexagon}
In this section, we describe a method to check the BJW conjecture (\ref{eq:PT}) explicitly for a few magnons. For simplicity, we consider the case where the excitations for the heavy operators are the transverse scalars $\Phi_{1\dot{1}},{\Phi}_{2\dot{2}}$ and the light operator being the BPS operator $\mathcal{O}_3=\tr\tilde{Z}^{2l_0}$. We will check the BJW conjecture explicitly for $N=1$ and $N=2$. The method can be readily applied to more general cases.

\subsection{Diagonal limit and kinematical poles}
\begin{figure}[h!]
\begin{center}
\includegraphics[scale=0.5]{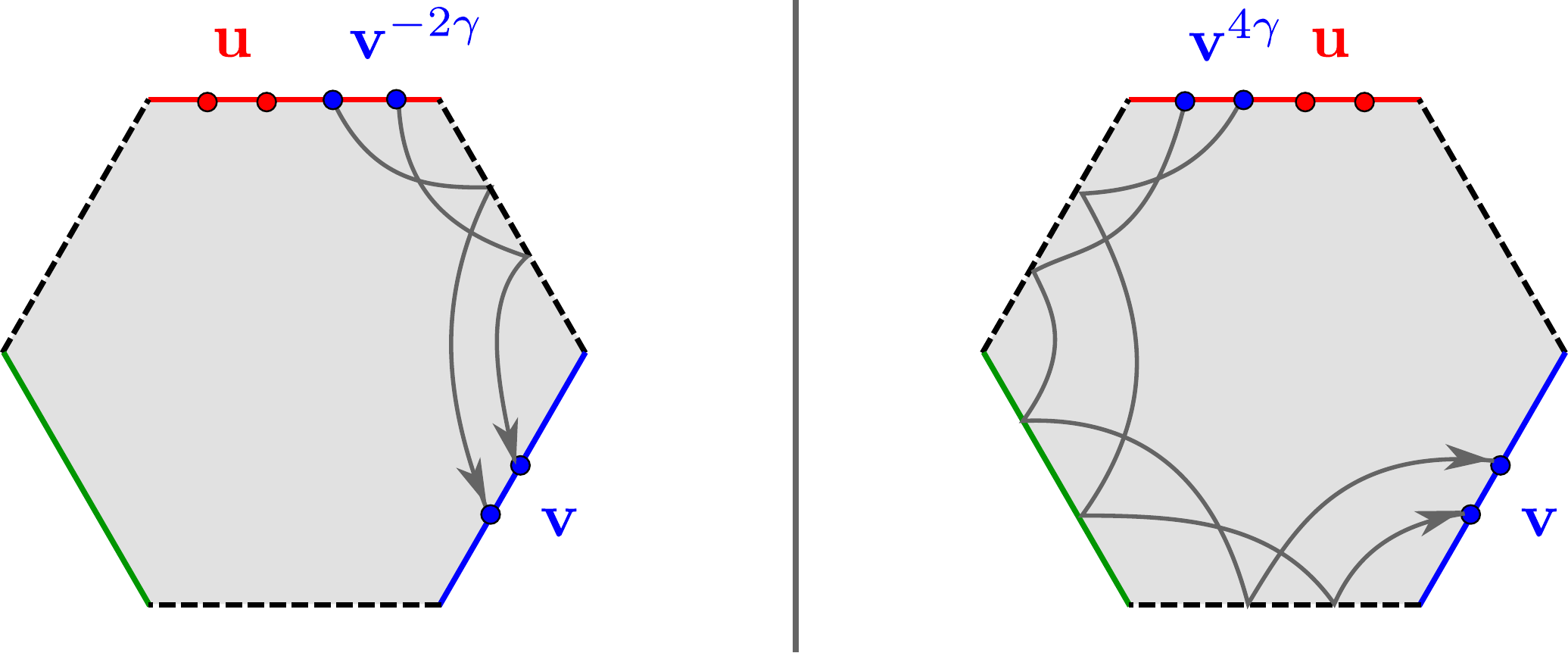}
\caption{Two possible crossing transformations. The left diagram corresponds to a $-2\gamma$ transformation and the right diagram corresponds to a $4\gamma$ transformation}
\label{fig:2r4r}
\end{center}
\end{figure}

In order to calculate the hexagon form factors, one needs to use mirror/crossing transformations to move all the excitations on the same edge. In our current example, we choose to move all the excitations on the edge which corresponds to $\mathcal{O}_1$. There are two possible transformations, as is shown in figure\,\ref{fig:2r4r}.
The two crossing transformations lead to the same final result, as is should be. However, the intermediate steps are rather different. In the diagonal limit where $u_i\to v_i$, there is a kinematical pole in the hexagon form factor. The hexagon form factor can be written as the product of a scalar or dynamical part and a matrix part. If we perform the $-2\gamma$ transformation, the kinematical pole appears in the matrix part while if we perform the $4\gamma$ transformation, the kinematical pole appears in the dynamical part. Since the dynamical part is a simple product of the scalar functions $h(u,v)$, it is much easier to keep track of the kinematical poles. At the same time, when performing crossing transformations, there will be some phase factors which originate from changing between the string frame and the spin chain frame. In the $4\gamma$ transformation, this phase factor is usually simpler. For the current case, it is simply 1. Therefore, we will proceed our calculation by performing $4\gamma$ transformations for excitations of $\mathcal{O}_2$.\par

The general hexagon form factor in our example takes the form
\begin{align}
\rH(\mathbf{u}|\mathbf{v})=\texttt{phase}_{4\gamma}\,\rH(\mathbf{v}^{4\gamma};\mathbf{u})
\end{align}
where $\texttt{phase}_{4\gamma}=1$ is the phase factor alluded before. Here and after, the hexagon with excitations on different edges is denoted as $\rH(\mathbf{u}|\mathbf{w}|\mathbf{v})$ while the one with excitations on the same edge is denoted by $\rH(\mathbf{u};\mathbf{w};\mathbf{v})$. The latter is called the \textit{fundamental hexagon} and can be written as a product of the dynamical part and matrix part\footnote{In general there is also a phase factor $(-1)^{\mathfrak{f}}$ taking into account the proper grading. For scalar excitations, this phase factor is simply 1.}
\begin{align}
\rH(\mathbf{v}^{4\gamma},\mathbf{u})=\rH^{\text{dyn}}(\mathbf{v}^{4\gamma},\mathbf{u})\rH^{\text{mat}}(\mathbf{v}^{4\gamma},\mathbf{u})
\end{align}
where
\begin{align}
\rH^{\text{dyn}}(\mathbf{v}^{4\gamma},\mathbf{u})=\frac{h^<(\mathbf{v},\mathbf{v})h^>(\mathbf{u},\mathbf{u})}{h(\mathbf{u},\mathbf{v})}.
\end{align}
Here we have introduced the short-hand notation
\begin{align}
F^<(\mathbf{v},\mathbf{v})=\prod_{i<j}F(v_i,v_j),\quad F^>(\mathbf{u},\mathbf{u})=\prod_{i>j}F(u_i,u_j),\quad F(\mathbf{u},\mathbf{v})=\prod_{i,j}F(u_i,v_j)
\end{align}
for any function $F(u,v)$ and have used the property $h(v^{4\gamma},u)=1/h(u,v)$. The scalar function $h(u,v)$ can be written as
\begin{align}
h(u,v)=\frac{u-v}{u-v-i}\tilde{h}(u,v),\qquad \tilde{h}(u,v)=\frac{(1-1/x_1^-x_2^+)^2}{(1-1/x_1^-x_2^-)(1-1/x_1^+x_2^+)}\frac{1}{\sigma_{12}}
\end{align}
where $x^\pm_1=x(u\pm i/2)$ and $x^\pm_2=x(v\pm i/2)$ are the Zhukowsky variables satisfying $x+1/x=u/g$ and $\sigma_{12}$ is the square root of BES dressing phase \cite{Beisert:2006ez}. The scalar function $\tilde{h}(u,u)$ is related to the measure $\mu(u)$ as
\begin{align}
\mu(u)=\frac{1}{\tilde{h}(u,u)}.
\end{align}
It it clear that in the diagonal limit $v\to u$, there is a kinematical pole in the function
\begin{align}
\frac{1}{h(u,v)}=\left(1-\frac{i}{u-v}\right)\frac{1}{\tilde{h}(u,v)}
\end{align}
as expected. We can thus write the dynamical part as
\begin{align}
\label{eq:Hd}
\rH^{\text{dyn}}(\mathbf{v}^{4\gamma},\mathbf{u})=\left(1-\frac{i}{\mathbf{u}-\mathbf{v}}\right)
\frac{h^<(\mathbf{v},\mathbf{v})h^>(\mathbf{u},\mathbf{u})}{\tilde{h}(\mathbf{u},\mathbf{v})}
\end{align}
where the kinematical poles in the diagonal limit are all in the first factor of (\ref{eq:Hd}).\par

The matrix part of the hexagon is given in terms of Beisert's $S$-matrix elements \cite{Beisert:2006qh} with the dressing phase setting to 1. Under $4\gamma$ transformation, the Zhukowsky variables are invariant $x^\pm(u^{4\gamma})=x^\pm(u)$ and hence the $S$-matrix elements and the matrix part of the hexagon form factor are also invariant
\begin{align}
\rH^{\text{mat}}(\mathbf{v}^{4\gamma},\mathbf{u})=\rH^{\text{mat}}(\mathbf{v},\mathbf{u}).
\end{align}

\subsection{$N=1$ case}
We first consider the simplest case, $\mathbf{u}=\{u_1\}$ and $\mathbf{v}=\{v_1\}$. There are two terms in the sum-over-partition formula
\begin{align}
C_{123}^{\bullet\bullet\circ}(u_1;v_1)=\texttt{t}_1+\texttt{t}_2
\end{align}
where
\begin{align}
\texttt{t}_1=&\,\left(1+\frac{i}{u_1-v_1}\right)\frac{\rH^{\text{mat}}(u_1,v_1)}{\tilde{h}(v_1,u_1)}\\\nonumber
\texttt{t}_2=&\,\left(1-\frac{i}{u_1-v_1}\right)\frac{\rH^{\text{mat}}(v_1,u_1)}{\tilde{h}(u_1,v_1)}\times e^{-i(p(u_1)-p(v_1))l}
\end{align}
In the diagonal limit, we can take $v_1=u_1-\epsilon$ and $\epsilon\to 0$. The sum $\texttt{t}_1+\texttt{t}_2$ can be rearranged as
\begin{align}
\texttt{t}_1+\texttt{t}_2=\rT_0+\frac{i}{\epsilon}\rT_1
\end{align}
where
\begin{align}
\rT_0=&\,\frac{\rH(u_1,v_1)}{\tilde{h}(v_1,u_1)}+\frac{\rH(v_1,u_1)}{\tilde{h}(u_1,v_1)}\times e^{-i(p(u_1)-p(v_1))l},\\\nonumber
\rT_1=&\,\frac{\rH(u_1,v_1)}{\tilde{h}(v_1,u_1)}-\frac{\rH(v_1,u_1)}{\tilde{h}(u_1,v_1)}\times e^{-i(p(u_1)-p(v_1))l}.
\end{align}
Here and later in this section, we omit the upper index of $\rH^{\text{mat}}$ to simplify the notation. The next step is to expand each term $\rT_0$ and $\rT_1$ in terms $\epsilon$ and keep only the leading term. The diagonal limit of the finite volume form factor is well defined, so we should have
\begin{align}
\rT_0=&\,\rT_{0,0}+\epsilon\,\rT_{0,1}+\cdots\\\nonumber
\rT_1=&\,\epsilon\,\rT_{1,1}+\epsilon^2\,\rT_{1,2}+\cdots
\end{align}
Namely, the $\epsilon$ expansion of $\rT_1$ starts at order $\mathcal{O}(\epsilon)$ and $\rT_{1,0}=0$. This fact can be seen easily since $\left.\rT_1\right|_{\epsilon=0}=0$ automatically. This is special for the one magnon case. We will see in the next subsection that for more magnons, the fact that the diagonal form factors are well-defined is ensured by the factorization properties of the hexagon. The un-normalized structure constant reads
\begin{align}
\mathcal{C}_{\text{HHL}}(u_1)=\rT_{0,0}+\rT_{1,1}=-\mu(u_1)\left(\rho_1(u_1)+F^s_1(u_1)\right),
\end{align}
where
\begin{align}
\label{eq:f1}
\rho_1(u)=&\,L\,p'(u),\\\nonumber
F^s_1(u)=&\,i \rH^{(0,1)}(u,u)-i \rH^{(1,0)}(u,u)+\frac{i \tilde{h}^{(0,1)}(u,u)}{\tilde{h}(u,u)}-\frac{i\tilde{h}^{(1,0)}(u,u)}{\tilde{h}(u,u)}-l_0 p'(u)-2
\end{align}
We see indeed that the volume dependence is encoded in the function $\rho_1(u)$. The infinite volume form factor for one magnon is given by $F^s_1(u)$. Here the upper indices $(1,0)$ and $(0,1)$ denote partial derivatives. For example,
\begin{align}
\rH^{(1,0)}(u,u)=\left.\frac{\partial}{\partial v}\rH(v,u)\right|_{v=u}.
\end{align}
We confirm that the normalized structure constant for one excitation indeed takes the form
\begin{align}
C_{\text{HHL}}(u_1)=\frac{1}{\rho_1(u_1)}\left(\rho_1(u_1)+F^s_1(u_1)\right)
\end{align}

\subsection{$N=2$ case}
For two magnon case, there are 6 terms
\begin{align}
C_{123}^{\bullet\bullet\circ}(u_1,u_2;v_1,v_2)=\sum_{i=1}^6 \texttt{t}_i.
\end{align}
where
\begin{align}
\texttt{t}_1=&\,\left(1+\frac{i}{u_1-v_1}\right)\left(1+\frac{i}{u_2-v_2}\right)\frac{h(u_2,u_1)h(v_1,v_2)}{h(v_1,u_2)h(v_2,u_1)}\,\frac{\rH(u_2,u_1,v_1,v_2)}{\tilde{h}(v_1,u_1)\tilde{h}(v_2,u_2)}\\\nonumber
\texttt{t}_2=&\,\left(1+\frac{i}{u_1-v_1}\right)\left(1-\frac{i}{u_2-v_2}\right)\frac{\rH(u_1,v_1)}{\tilde{h}(v_1,u_1)}\frac{\rH(v_2,u_2)}{\tilde{h}(u_2,v_2)}\times e^{-i(p(u_2)-p(v_2))l}\\\nonumber
\texttt{t}_3=&\,\frac{\rH(u_1,v_2)}{h(v_2,u_1)}\frac{\rH(v_1,u_2)}{h(u_2,v_1)}\times e^{-i(p(u_2)-p(v_1))l}\,S(v_1,v_2)\\\nonumber
\texttt{t}_4=&\,\frac{\rH(u_2,v_1)}{h(v_1,u_2)}\frac{\rH(v_2,u_1)}{h(u_1,v_2)}\times e^{-i(p(u_1)-p(v_2))l}\,S(u_2,u_1)\\\nonumber
\texttt{t}_5=&\,\left(1-\frac{i}{u_1-v_1}\right)\left(1+\frac{i}{u_2-v_2}\right)\frac{\rH(u_2,v_2)}{\tilde{h}(v_2,u_2)}\frac{\rH(v_1,u_1)}{\tilde{h}(u_1,v_1)}\times e^{-i(p(u_1)-p(v_1))l}S(u_2,u_1)S(v_1,v_2)\\\nonumber
\texttt{t}_6=&\,\left(1-\frac{i}{u_1-v_1}\right)\left(1-\frac{i}{u_2-v_2}\right)\frac{h(u_2,u_1)h(v_1,v_2)}{h(u_1,v_2)h(u_2,v_1)}\frac{\rH(v_1,v_2,u_2,u_1)}{\tilde{h}(u_1,v_1)\tilde{h}(u_2,v_2)}
\times e^{-i(p(u_1)+p(u_2)-p(v_1)-p(v_2))}.
\end{align}
In the diagonal limit, we take $v_k=u_k-\epsilon$ and arrange the sum as
\begin{align}
\sum_{i=1}^6\texttt{t}_i=\rT_0+\frac{i}{\epsilon}\rT_1+\frac{i^2}{\epsilon^2}\rT_2.
\end{align}
Then we perform the $\epsilon$ expansion for each term, $\rT_k=\sum_{n=0}^\infty\rT_{k,n}\epsilon^n$, $k=0,1,2$. We should have $\rT_{1,0}=\rT_{2,0}=\rT_{2,1}=0$ in order the diagonal limit to be well-defined. The un-normalized diagonal structure constant is given by
\begin{align}
\mathcal{C}_{\text{HHL}}(u_1,u_2)=\rT_{0,0}+\rT_{1,1}+\rT_{2,2}.
\end{align}
As alluded before, the disappearance of $\rT_{k,n}$ $(n<k)$ is guaranteed by the factorization property of the hexagon form factors. For example,
\begin{align}
\rT_{2,0}=\frac{1}{\tilde{h}(u_1,u_1)\tilde{h}(u_2,u_2)}\left[\rH(u_1,u_2,u_2,u_1)+\rH(u_2,u_1,u_1,u_2)-2\rH(u_1,u_1)\rH(u_2,u_2)  \right]
\end{align}
which does not vanish automatically. However, notice that there are coinciding rapidities in the matrix part of the hexagon, they can be written in terms of hexagons with less excitations. In fact, we will derive in the next section that
\begin{align}
\label{eq:fac}
\rH(u_1,u_2,u_2,u_1)=\rH(u_1,u_2,u_2,u_1)=\rH(u_1,u_1)\rH(u_2,u_2).
\end{align}
Taking into account (\ref{eq:fac}), we have indeed $\rT_{2,0}=0$. Similarly, $\rT_{2,1}$ does not vanish automatically, but will vanish if we take into account (\ref{eq:fac}) as well as the relations of the following type
\begin{align}
\label{eq:dH}
\rH^{(1,0,0,0)}(u_1,u_2,u_2,u_1)=&\,\rH^{(1,0)}(u_1,u_1)+\frac{h^{(0,1)}(u_2,u_1)}{h(u_2,u_1)}-\frac{h^{(1,0)}(u_2,u_1)}{h(u_2,u_1)}+\frac{S^{(1,0)}(u_1,u_2)}{S(u_1,u_2)}
\end{align}
Here $\rH^{(1,0,0,0)}(u_1,u_2,u_2,u_1)$ stands for
\begin{align}
\rH^{(1,0,0,0)}(u_1,u_2,u_2,u_1)\equiv\left.\frac{\partial}{\partial v}\rH(v,u_2,u_1,u_2)\right|_{v=u_1}
\end{align}
The relation (\ref{eq:dH}) comes from taking the derivatives with respect to $v$ of the following factorization relation
\begin{align}
\rH(v,u_2,u_2,u_1)=&\,S(v_,u_2)S(u_2,u_1)\rH(v,u_1)
\end{align}
The mechanism works also for more magnons. By using the factorization properties and the corresponding derivatives, the terms $\rT_{k,n}$ with $n<k$ will vanish. Taking into account the normalization, we find that the normalized symmetric structure constant takes the following form
\begin{align}
{C}_{\text{HHL}}(u_1,u_2)=\frac{1}{\rho_2(u_1,u_2)}\left[\rho_2(u_1,u_2)+\rho_1^s(u_1)F^s_1(u_2)+\rho_1^s(u_2)F^s_1(u_1)+F^s_2(u_1,u_2)\right]
\end{align}
where $F^s_1(u)$ is derived in (\ref{eq:f1}) and $F^s_2(u_1,u_2)$ a rather complicated function in terms of the momenta $p(u)$, $\mathfrak{su}(2)$ scattering matrix $S(u,v)$, the scalar factor $h(u,v)$, the matrix part of the hexagon for 2 and 4 excitations $\rH(u_1,v_1)$, $\rH(u_1,u_2,v_1,v_2)$ and their derivatives. The explicit form of $F_2^s(u_1,u_2)$ can be found in appendix\,\ref{sec:F2}.

\subsection{Generalization to $N$ magnons}
The generalization to $N$ magnon case is now straightforward. In the diagonal limit, $v_k=u_k-\epsilon$ with $\epsilon\to 0$, we can organize the result as
\begin{align}
\sum_{k=0}^N \frac{i^k}{\epsilon^k}\rT_k.
\end{align}
Then we expand each $\rT_k$ in terms of $\epsilon$
\begin{align}
\rT_k=\sum_{n=0}^\infty \rT_{k,n}\,\epsilon^n.
\end{align}
The un-normalized symmetric structure constant is given by
\begin{align}
\mathcal{C}_{\text{HHL}}(\mathbf{u})=\sum_{k=1}^N \rT_{k,k}.
\end{align}
In order to check (\ref{eq:PT}) for $N$ magnons, we need to know the expression of all the \emph{infinite volume form factors} $F_n^s(u_1,\cdots,u_n)$ with $n<N$ in terms of $p(u),h(u,v),S(u,v)$ and $\rH$. Then by subtracting the volume dependence from the finite volume form factor of $N$ magnons, we obtain the infinite volume form factor of $N$ magnons $F^s_N(u_1,\cdots,u_N)$.\par

We can check the structure (\ref{eq:PT}) for a few excitations. The expression for the infinite volume form factors become complicated very quickly. Although a general proof is very hard to achieve following this method, we can give an argument for (\ref{eq:PT}) based on our calculation.

In our previous calculations for one and two excitations, we do not specify the explicit form of $p(u),h(u,v),S(u,v)$ and $\rH$. The calculation is exactly the same whether we take the leading order expressions or the all-loop expressions. The only differences are the explicit form of $\rho^s$ and infinite volume form factors $F^s$. As far as the structure (\ref{eq:PT}) is concerned, they are equivalent. If we can find a ``representation'' of the quantities $p(u)$, $h(u,v)$, $S(u,v)$ and $\rH$ such that the structure (\ref{eq:PT}) holds, then the BJW conjecture should hold in general. In our case, such a ``representation'' indeed exists, where we take all the quantities $p,h,S,\rH$ at the leading order. In \cite{Hollo:2015cda}, we have shown that (\ref{eq:PT}) holds for any magnons at the leading order using the solution of the quantum inverse scattering problem and the Slavnov determinant formula. Based on this argument and the explicit calculations of the first few magnons, we already see that the structure (\ref{eq:PT}) should hold at finite coupling\footnote{Again up to mirror excitations.}. The rigorous proof will be given in section\,\ref{sec:proof}.

\section{Factorization property}
\label{sec:factorization}
In this section, we derive the factorization property of the hexagon form factor. These properties are used in the previous section to insure the diagonal form factors to be well-defined and will be used in the next section to prove the BJW conjecture. The main result is
\begin{align}
\rH^{\text{mat}}_{\chi}\left(u,\mathbf{u},u\right)=(-1)^{\mathfrak{f}}\,n_\chi
\rH^{\text{mat}}_{\chi}\left(\mathbf{u}\right)
\end{align}
where the polarizations of the excitations are $(\chi(u),\mathcal{X}_{A_1\dot{A}_1}(u_1),\cdots,\mathcal{X}_{A_N\dot{A}_N}(u_N),\bar{\chi}(u))$, with $\mathcal{X}_{\mathbf{A\dot{A}}}(\mathbf{u})=\mathcal{X}_{A_1\dot{A}_1}(u_1)\cdots\mathcal{X}_{A_N\dot{A}_N}(u_N)$ being arbitrary. The phase factor $(-1)^{\mathfrak{f}}$ takes into account the proper grading and is given in (\ref{eq:deff}) and (\ref{eq:ff}) and $n_\chi$ is a simple number define in (\ref{eq:nchi}) and calculated in appendix \ref{sec:nchi}.\par

\begin{figure}[h!]
\begin{center}
\includegraphics[scale=0.5]{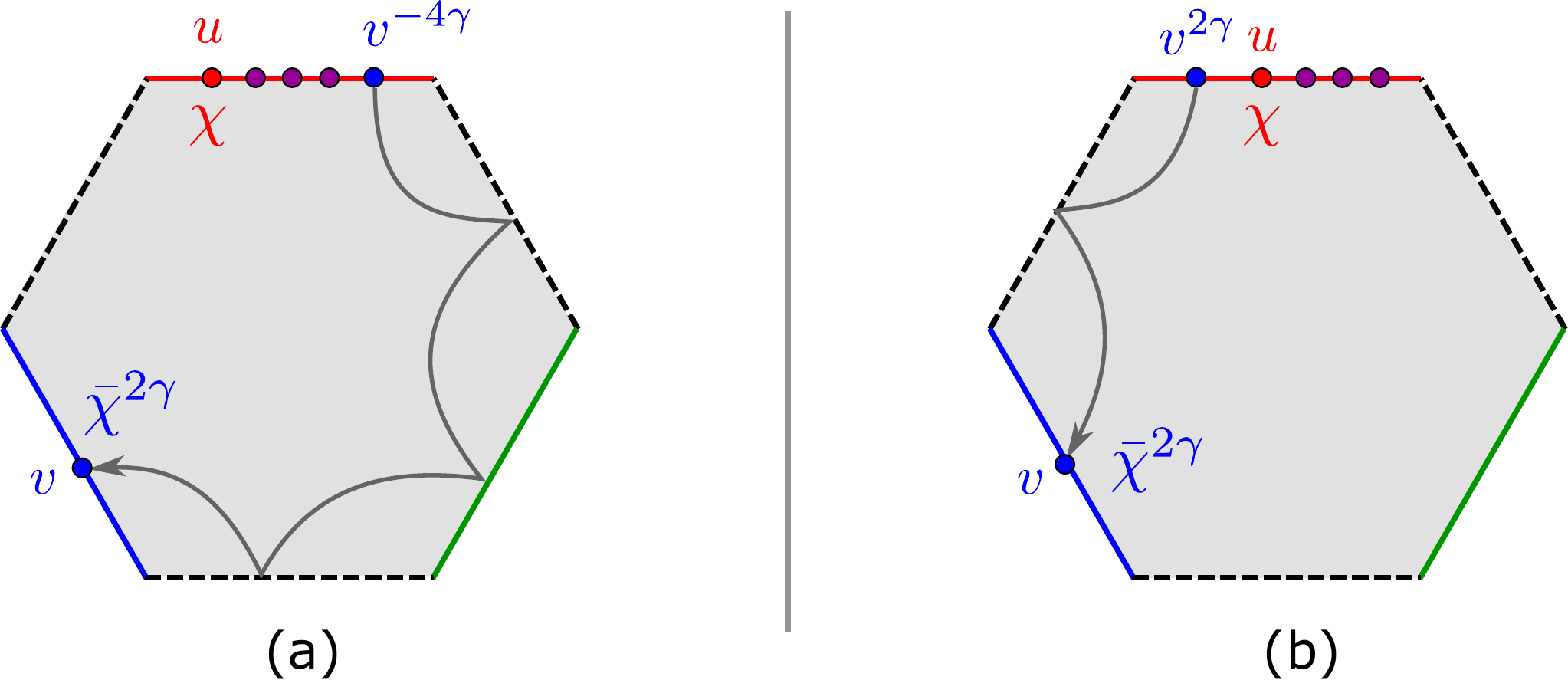}
\caption{The two different transformations: (a) the $-4\gamma$ transformation; (b) the $2\gamma$ transformation.}
\label{fig:crossing}
\end{center}
\end{figure}

To prove the factorization properties, we compute the following hexagon form factor
\begin{align}
\rH_{\chi}=\langle\mathfrak{h}|\chi(u)\mathcal{X}_{\mathbf{A}\dot{\mathbf{A}}}(\mathbf{u})\rangle|\bar{\chi}^{2\gamma}(v)\rangle|0\rangle.
\end{align}
in the limit $v\to u$. We compute the hexagon by performing crossing transformations for $\bar{\chi}$. We can choose either a $2\gamma$ transformation or a $-4\gamma$ transformation, as is shown in figure\,\ref{fig:crossing}.
By comparing the expressions for two different crossing transformations, we obtain the factorization properties. The two mirror transformations should lead to the same result
\begin{align}
\label{eq:simple}
\rH_{\chi}=\texttt{phase}_{2\gamma}^\chi\,\rH_{\chi,2\gamma}=\texttt{phase}_{-4\gamma}^\chi\,\rH_{\chi,-4\gamma}
\end{align}
where $\texttt{phase}_{2\gamma}^\chi$, $\texttt{phase}_{-4\gamma}^\chi$ are the phase factors coming from the crossing transformations and $\rH_{\chi,2\gamma}$, $\rH_{\chi,-4\gamma}$ are the corresponding fundamental hexagons. From (\ref{eq:simple}), we have
\begin{align}
\label{eq:2g4g}
\rH_{\chi,-4\gamma}=\frac{\texttt{phase}_{2\gamma}^\chi}{\texttt{phase}_{-4\gamma}^\chi}\,\rH_{\chi,2\gamma}.
\end{align}
The fundamental hexagons can be written as a product of  the dynamical part, the matrix part and the phase factor which takes into account the grading. Let us denote the ratio of the two phase factors by $(-1)^{\mathfrak{f}}$, namely
\begin{align}
\label{eq:deff}
\mathfrak{f}=\mathfrak{f}_{2\gamma}-\mathfrak{f}_{-4\gamma}
\end{align}
where
\begin{align}
\label{eq:ff}
\mathfrak{f}_{2\gamma}=&\,(\dot{f}_{\bar{\chi}}+\dot{f}_{\chi})f_{\mathbf{A}}+\dot{f}_{\bar{\chi}}f_\chi,\\\nonumber
\mathfrak{f}_{-4\gamma}=&\,\dot{f}_{\chi}f_{\mathbf{A}}+\dot{f}_{\dot{\mathbf{A}}}f_{\bar{\chi}^{2\gamma}}+\dot{f}_{\chi}f_{\bar{\chi}^{2\gamma}}.
\end{align}
Here the symbol $f$ and $\dot{f}$ denote the fermionic number for the corresponding excitation of the undotted and dotted indices and $f_{\mathbf{A}}=\sum_{i=1}^n f_{A_i}$, $\dot{f}_{\dot{\mathbf{A}}}=\sum_{i=1}^n \dot{f}_{\dot{A}_i}$. Let us notice that for the 8 pairs of excitations $\chi,\bar{\chi}$, we always have $\dot{f}_\chi+\dot{f}_{\bar{\chi}}\equiv 0$ (mod 2).
Therefore $\mathfrak{f}_{2\gamma}\equiv\dot{f}_{\bar{\chi}}f_\chi$ (mod 2). For the l.h.s. of (\ref{eq:2g4g}),
\begin{align}
\rH_{\chi,-4\gamma}^{\text{dyn}}=\frac{1}{h(v,u)}\frac{h(u,\mathbf{u})}{h(v,\mathbf{u})},
\quad\rH_{\chi,-4\gamma}^{\text{mat}}=\rH_{\chi,-4\gamma}^{\text{mat}}(u,\mathbf{u},v).
\end{align}
In the limit $v\to u$, there is a kinematical pole in the dynamical part while the matrix part is regular
\begin{align}
\label{eq:4gd}
\underset{v\to u}{\text{Res}}\,\rH_{\chi,-4\gamma}^{\text{dyn}}=&\,i\mu(u),\\\nonumber
\lim_{v\to u}\rH_{\chi,-4\gamma}^{\text{mat}}=&\,\rH_{\chi,-4\gamma}^{\text{mat}}(u,\mathbf{u},u).
\end{align}
For the r.h.s. of (\ref{eq:2g4g}),
\begin{align}
\rH_{\chi,2\gamma}^{\text{dyn}}=h(v^{2\gamma},\mathbf{u})h(u,\mathbf{u}),\quad \rH_{\chi,2\gamma}^{\text{mat}}=\rH_{\chi,2\gamma}^{\text{mat}}(v^{2\gamma},u,\mathbf{u}).
\end{align}
In the decoupling limit $v\to u$, the dynamical part is regular while the matrix part has a pole. The residue of the matrix part can be worked out by using the same argument as in \cite{Basso:2015zoa}
\begin{align}
\label{eq:2gd}
\lim_{v\to u}\rH_{\chi,2\gamma}^{\text{dyn}}=&\,h(u^{2\gamma},\mathbf{u})h(u,\mathbf{u})\\\nonumber
\underset{v\to u}{\text{Res}}\,\rH_{\chi,2\gamma}^{\text{mat}}=&\,\underset{v\to u}{\text{Res}}\,\rH_{\chi,2\gamma}^{\text{mat}}(v^{2\gamma},u)\cdot
\frac{\rH^{\text{mat}}(\mathbf{u})}{h(u^{2\gamma},\mathbf{u})h(u,\mathbf{u})}\cdot e^{ic_\chi P}.
\end{align}
Here $c_\chi=1,0,\frac{1}{2}$ for scalars, derivatives and fermions. $P$ is the total momentum of the excitations $\mathcal{X}_{\mathbf{A}\dot{\mathbf{A}}}(\mathbf{u})$
\begin{align}
P=\sum_{i=1}^N p(u_i)
\end{align}
The derivation of (\ref{eq:2gd}) can be found in appendix\,\ref{sec:singlet}. The ratio of the phase factors $\texttt{phase}_{2\gamma}^\chi$ and $\texttt{phase}_{-4\gamma}^\chi$ is computed in appendix\,\ref{sec:ratio} and reads
\begin{align}
\label{eq:ratio}
\frac{\texttt{phase}_{2\gamma}^\chi}{\texttt{phase}_{-4\gamma}^\chi}=-e^{-ic_\chi(p+P)},\qquad p=p(u).
\end{align}
Combining (\ref{eq:deff}), (\ref{eq:4gd}), (\ref{eq:2gd}) and (\ref{eq:ratio}), we have
\begin{align}
\rH_{\chi,-4\gamma}^{\text{mat}}(u,\mathbf{u},u)=(-1)^{\mathfrak{f}}\,\rH_{\chi,2\gamma}^{\text{mat}}(\mathbf{u})\cdot n_\chi
\end{align}
where
\begin{align}
\label{eq:nchi}
n_\chi=\frac{i\,\underset{v\to u}{\text{Res}}\,\rH_{\chi,2\gamma}^{\text{mat}}(v^{2\gamma},u)}{\mu(u)\,e^{ic_\chi p}}
\end{align}
is a simple number and is computed in appendix\,\ref{sec:nchi}. We list $n_\chi$ for the 8 pairs of excitations in the following table
\begin{center}
\begin{tabular}{|c|c|c|c|c|c|c|c|c|}
  \hline
  $\chi$       & $\Phi_{1\dot{1}}$            & $\Phi_{1\dot{2}}$            & $\mathcal{D}_{3\dot{3}}$            &$\mathcal{D}_{3\dot{4}}$  & $\Psi_{1\dot{3}}$ & $\Psi_{2\dot{3}}$ & $\Psi_{1\dot{4}}$ & $\Psi_{2\dot{4}}$ \\
  \hline
  $\bar{\chi}$ & ${\Phi}_{2\dot{2}}$ & ${\Phi}_{2\dot{1}}$ & ${\mathcal{D}}_{4\dot{4}}$ &${\mathcal{D}}_{4\dot{3}}$ & ${\Psi}_{2\dot{4}}$ & ${\Psi}_{1\dot{4}}$ & ${\Psi}_{2\dot{3}}$ & ${\Psi}_{1\dot{3}}$ \\
  \hline
  $n_{\chi}$   & $1 $                         & $-1$                         & $1$                                 & $-1$ & $1$ & $-1$ & $-1$ & $1$ \\
  \hline
\end{tabular}
\end{center}
For our purpose, we are concerned with the following type of factorization property
\begin{align}
\label{eq:fac}
\rH^{\text{mat}}(u,\mathbf{u};\mathbf{w};\mathbf{v},u)=(-1)^{\mathfrak{f}}\,n_\chi\,\rH^{\text{mat}}(\mathbf{u};\mathbf{w};\mathbf{v})
\end{align}
where $\{u,\mathbf{u}\}$ and $\{\mathbf{v},u\}$ are rapidities of the excitations of type $\chi$ and $\bar{\chi}^{2\gamma}$, respectively. The polarizations of $\mathbf{w}$ can be arbitrary. If the coinciding rapidities are not on the leftmost and rightmost, we can use the following relation to move the excitations
\begin{align}
\label{eq:perm}
\rH^{\text{mat}}(\cdots,u_i,u_j,\cdots;\star;\star)=&\,S_\chi(u_i,u_j)\frac{h(u_j,u_i)}{h(u_i,u_j)}\rH^{\text{mat}}(\cdots,u_j,u_i,\cdots;\star;\star)\\\nonumber
\rH^{\text{mat}}(\star;\star;\cdots,u_i,u_j,\cdots)=&\,S_\chi(u_i,u_j)\frac{h(u_j,u_i)}{h(u_i,u_j)}\rH^{\text{mat}}(\star;\star;\cdots,u_j,u_i,\cdots).
\end{align}
Equation (\ref{eq:fac}) and (\ref{eq:perm}) together give the factorization property.

\section{Proof of BJW conjecture}
\label{sec:proof}
In this section, we prove the BJW conjecture up to mirror excitations. We first prove a recursion relation for the un-normalized HHL structure constant and then prove the BJW conjecture based on the recursion relation. This is a generalization of the proof presented in \cite{Jiang:2015bvm}.

\subsection{The recursion relation}
As we can see from the examples, the explicit $L$-dependence comes from taking derivatives of the phase factor $\xi(v_i)=e^{ilp(v_i)}$. This implies that the polynomial dependence of $l$ always enters through the combination $z_i=lp'(u_i)$. It proves to be useful to consider the $z_i$-dependence of the structure constant. Let us first introduce some notations. We denote the expression in the sum-over-partition formula (\ref{eq:CSoP}) by $\mathcal{C}_{2N}^{\mathbf{w}}(\mathbf{u}|\mathbf{v})\equiv C_{123}^{\bullet\bullet\bullet}$. The diagonal limit of this quantity is denoted by
\begin{align}
\label{eq:C2N}
\mathcal{C}_{\text{HHL}}^{\mathbf{w}}(\mathbf{u})=\lim_{\mathbf{v}\to\mathbf{u}}\mathcal{C}_{2N}^{\mathbf{w}}(\mathbf{u}|\mathbf{v}).
\end{align}
Note that the sum-over-partition formula gives the structure constant in the large but finite volume, therefore the diagonal limit is unambiguously defined and does not depend on the way we take the diagonal limit. We can take $v_i=u_i-\epsilon_i$ and then take $\epsilon_i\to 0$ one by one, or equivalently we can take $\epsilon_i=\epsilon$ and take $\epsilon\to0$, they give the same result. This is different from the diagonal limit in \textit{infinite volume} where the result is divergent and depends on how one takes the diagonal limit. Another useful quantity in the diagonal limit is given by
\begin{align}
\label{eq:FN}
\mathcal{F}^{\mathbf{w}}_N(\mathbf{u})=\lim_{\mathbf{v}\to\mathbf{u}}\left(\left.\mathcal{C}_{2N}^{\mathbf{w}}(\mathbf{u}|\mathbf{v})\right|_{\xi(v_i)=\xi(u_i)}\right)
\end{align}
In terms of words, we first put the phase factor $e^{ilp(v_i)}\to e^{ilp(u_i)}$ and then take the diagonal limit. As we discussed before, the explicit $l$-dependence originates from derivatives of the factor $\xi(v_i)$. Replacing these factors by $\xi(u_i)$ before taking the diagonal limit eliminates the $l$-dependence. Therefore $\mathcal{F}^{\mathbf{w}}_N(\mathbf{u})$ does not depend on $l$ and is a well defined quantity in the infinite volume. In both (\ref{eq:C2N}) and (\ref{eq:FN}), after taking the diagonal limit, we impose the BAE to replace the phase factors $e^{-ilp(u_i)}$ by $e^{il_0p(u_i)}$ together with products of $S$-matrices.\par

The dependence of $\mathcal{C}_{\text{HHL}}^{\mathbf{w}}(\mathbf{u})$ on $z_k$ is linear and is given by the following relation
\begin{align}
\label{eq:recur}
\frac{\partial}{\partial z_k}\mathcal{C}_{\text{HHL}}^{\mathbf{w}}(\mathbf{u})=a_{\chi,k}\,\mathcal{C}_{\text{HHL}}^{\mathbf{w},\text{mod}}(\mathbf{u}\setminus u_k),\quad k=1,\cdots,N.
\end{align}
where the set $\mathbf{u}\setminus u_k$ means the rapidity $u_k$ is deleted from the original set and
\begin{align}
a_{\chi,k}\equiv a_\chi(u_k)=(-1)^{\dot{f}_{\bar{\chi}}f_\chi+1}\,n_\chi\,\mu(u_k)
\end{align}
The index ``mod'' stands for the following replacement
\begin{align}
\label{eq:mod}
z_i\to z_i^{\text{mod}}=z_i+\varphi(u_i,u_k),\qquad \varphi(u,v)=-i\frac{\partial}{\partial u}\log S(u,v).
\end{align}

We first prove the recursion relation for $z_N$. The quantity $z_N$ comes from taking derivatives of the factor $e^{ip(v_N)l}$, therefore we must have $v_N\in\bar{\beta}$ in order to have such a factor. On the other hand, we also need to have $u_N\in\bar{\alpha}$ because otherwise $u_N$ and $v_N$ are on different hexagons and there is no kinematical pole and hence not necessary to take derivatives. Consider a generic such term in the sum-over-partition formula (\ref{eq:CSoP}) denoted by $t(\{u_N\}\cup\bar{\alpha},\bar{\beta}\cup\{v_N\},\bar{\delta})$. The splitting factors satisfy
\begin{align}
\label{eq:split}
\frac{\omega_{-l}(\alpha,\{u_N\}\cup\bar{\alpha})\omega_{l}(\beta,\bar{\beta}\cup \{v_N\})\omega_{l_0}(\delta,\bar{\delta})}{\omega_{-l}(\alpha,\bar{\alpha})\omega_{l}(\beta,\bar{\beta})\omega_{l_0}(\delta,\bar{\delta})}
=e^{-ilp(u_N)+ilp(v_N)}
\end{align}
The hexagon form factor that we are interested in takes the following form
\begin{align}
\rH(\bar{\beta},v_N|\bar{\delta}|u_N,\bar{\alpha})=\texttt{phase}_{4\gamma}\,\cdot\rH(\bar{\beta}^{4\gamma},v_N^{4\gamma};\bar{\delta}^{2\gamma};u_N,\bar{\alpha})
\end{align}
We want to study the relation between this hexagon form factor and the one without $u_N$ and $v_N$
\begin{align}
\rH(\bar{\beta}|\bar{\delta}|\bar{\alpha})=\texttt{phase}'_{4\gamma}\,\cdot\rH(\bar{\beta}^{4\gamma};\bar{\delta}^{2\gamma};\bar{\alpha})
\end{align}
One can prove that $\texttt{phase}_{4\gamma}=\texttt{phase}'_{4\gamma}$ in the limit $v_N\to u_N$. The fundamental hexagon form factor is the product of a phase factor $(-1)^{\mathfrak{f}}$, the dynamical part and the matrix part. Let us denote the ratio of the phase factors of the hexagons $\rH(\bar{\beta}^{4\gamma},v_N^{4\gamma};\bar{\delta}^{2\gamma};u_N,\bar{\alpha})$ and $\rH(\bar{\beta}^{4\gamma};\bar{\delta}^{2\gamma};\bar{\alpha})$ by $(-1)^{\Delta\mathfrak{f}}$. The dynamical parts of the fundamental hexagons satisfy
\begin{align}
\label{eq:dyn}
\frac{\rH^{\text{dyn}}(\bar{\beta}^{4\gamma},v_N^{4\gamma};\bar{\delta}^{2\gamma};u_N,\bar{\alpha})}{\rH^{\text{dyn}}(\bar{\beta}^{4\gamma};\bar{\delta}^{2\gamma};\bar{\alpha})}
=\frac{h(\bar{\beta},v_N)}{h(u_N,\bar{\beta})}\frac{h(u_N,\bar{\alpha})}{h(\bar{\alpha},v_N)}\frac{h(\bar{\delta}^{2\gamma},u_N)}{h(\bar{\delta}^{2\gamma},v_N)}
\cdot\frac{1}{h(u_N,v_N)}
\end{align}
The splitting factor and the dynamical part are universal in the sense that they do not depend on the polarizations of excitations. For the matrix part of the hexagon, we apply the factorization property
\begin{align}
\label{eq:mat}
\frac{\rH^{\text{mat}}(\bar{\beta}^{4\gamma},v_N^{4\gamma};\bar{\delta}^{2\gamma};u_N,\alpha)}{\rH^{\text{mat}}(\bar{\beta}^{4\gamma};\bar{\delta}^{2\gamma};\alpha)}
=(-1)^{\mathfrak{f}}\,n_\chi S_{\chi}(\bar{\beta},u_N)S_{\chi}(u_N,\bar{\alpha})\frac{h(u_N,\bar{\beta})}{h(\bar{\beta},u_N)}\frac{h(\bar{\alpha},u_N)}{h(u_N,\bar{\alpha})}+\mathcal{O}(\epsilon)
\end{align}
where $n_\chi=\pm 1$ depending on the polarizations. One can show straightforwardly that
\begin{align}
(-1)^{\Delta\mathfrak{f}+\mathfrak{f}}=(-1)^{\dot{f}_{\bar{\chi}}f_\chi}.
\end{align}
Combining (\ref{eq:split}),(\ref{eq:dyn}) and (\ref{eq:mat}) and summing over the partitions, we obtain
\begin{align}
\frac{\partial}{\partial z_N}\lim_{\epsilon_N\to 0}\left.\mathcal{C}_{2N}^{\mathbf{w}}(\mathbf{u}|\mathbf{v})\right|_{v_N=u_N-\epsilon_N}
=a_{\chi,N}\,\mathcal{C}_{2(N-1)}^{\mathbf{w},\text{mod}}(\mathbf{u}\setminus u_N|\mathbf{v}\setminus v_N).
\end{align}
where again the index ``mod'' stands for the replacement rule (\ref{eq:mod}). After taking $v_i\to u_i$ for the rest of the rapidities, we obtain
\begin{align}
\frac{\partial}{\partial z_N}\mathcal{C}_{\text{HHL}}^{\mathbf{w}}(\mathbf{u})=a_{\chi,N}\,\mathcal{C}_{\text{HHL}}^{\mathbf{w},\text{mod}}(\mathbf{u}\setminus u_N).
\end{align}
Finally let us notice that the structure constant is symmetric with respect to the rapidities, hence (\ref{eq:recur}) is valid for any $k$.

\subsection{Proof of BJW conjecture}
Now we are ready to prove to the BJW conjecture up to finite size corrections. For a given partition $\mathbf{u}=\alpha\cup\bar{\alpha}$, let us define
\begin{align}
K_N(\bar{\alpha})=\prod_{k=1}^Na_\chi(u_k)\,\rho_{|\bar{\alpha}|,l}^s(\bar{\alpha})
\end{align}
where $\rho_{|\bar{\alpha}|,l}^s(\bar{\alpha})$ indicates the fact that it is defined with respect to the length $l=L-l_0$. We further define a quantity
\begin{align}
\mathcal{W}_N^{\mathbf{w}}(\mathbf{u})=\sum_{\alpha\cup\bar{\alpha}=\mathbf{u}}\mathcal{F}^{\mathbf{w}}_{|\alpha|}(\alpha)K_N(\bar{\alpha}).
\end{align}
As a first step, we want to show
\begin{align}
\label{eq:WC}
\mathcal{W}_N^{\mathbf{w}}(\mathbf{u})=\mathcal{C}_{\text{HHL}}^{\mathbf{w}}(\mathbf{u}).
\end{align}
Noticing that
\begin{align}
\frac{\partial}{\partial z_k}\rho^s_{N,l}(\mathbf{u})=\rho^{s,\text{mod}}_{N-1,l}(\mathbf{u}\setminus u_k)
\end{align}
with the modification rule given in (\ref{eq:mod}), we can deduce the $z_k$ dependence of $\mathcal{W}_N^{\mathbf{w}}$
\begin{align}
\label{eq:Wrecur}
\frac{\partial}{\partial z_k}\mathcal{W}_N^{\mathbf{w}}(\mathbf{u})=a_{\chi,k}\,\mathcal{W}_{N-1}^{\mathbf{w},\text{mod}}(\mathbf{u}\setminus u_k).
\end{align}
We can prove (\ref{eq:WC}) by induction. The case $n=1$ can be verified by explicit computation. Assume that (\ref{eq:WC}) holds for $n\le N-1$, we need to prove that it is also true for $n=N$. From (\ref{eq:recur}) and (\ref{eq:Wrecur}) we find that the $z_i$ dependence of the two quantities are the same. It remains to show that the terms independent of $z_i$ is also the same. Putting $z_i\to 0$, all $\rho^s_{|\bar{\alpha}|,l}(\bar{\alpha})=0$ and hence
\begin{align}
\left.\mathcal{W}_N^{\mathbf{w}}(\mathbf{u})\right|_{z_i=0}=\mathcal{F}^{\mathbf{w}}_N(\mathbf{u}).
\end{align}
On the other hand, form the definition of $\mathcal{F}^{\mathbf{w}}_N(\mathbf{u})$ (\ref{eq:FN}), we first put $e^{ilp(v_i)}$ to $e^{ilp(u_i)}$ and then take the diagonal limit, which prevents the appearance of $z_i$ and thus
\begin{align}
\label{eq:CF}
\left.\mathcal{C}_{\text{HHL}}^{\mathbf{w}}(\mathbf{u})\right|_{z_i=0}=\mathcal{F}^{\mathbf{w}}_N(\mathbf{u}).
\end{align}
This proves (\ref{eq:WC}) and we have
\begin{align}
\mathcal{C}_{\text{HHL}}^{\mathbf{w}}(\mathbf{u})=\prod_{k=1}^N a_\chi(u_k)
\sum_{\alpha\cup\bar{\alpha}=\mathbf{u}}\mathcal{F}^{\mathbf{w}}_{|\alpha|}(\alpha)\rho^s_{|\bar{\alpha}|,l}(\bar{\alpha}).
\end{align}
Finally we go from length $l$ to length $L$, this can be done by the following relation
\begin{align}
\rho^s_{N,l_1+l_2}(\mathbf{u})=\sum_{\alpha\cup\bar{\alpha}=\mathbf{u}}\rho^s_{|\alpha|,l_1}(\alpha)\,\rho^s_{|\bar{\alpha}|,l_2}(\bar{\alpha}).
\end{align}
Taking $l_1=L$ and $l_2=-l_0$, we have
\begin{align}
\mathcal{C}_{\text{HHL}}^{\mathbf{w}}(\mathbf{u})=\prod_{k=1}^N a_\chi(u_k)
\sum_{\alpha\cup\bar{\alpha}=\mathbf{u}}F^{\mathbf{w},s}_{|\alpha|}(\alpha)\rho^s_{|\bar{\alpha}|,L}(\bar{\alpha}).
\end{align}
where
\begin{align}
\label{eq:CFF}
F^{\mathbf{w},s}_{|\alpha|}(\alpha)=\sum_{\beta\cup\bar{\beta}=\alpha}\mathcal{F}^{\mathbf{w}}_{|\beta|}(\beta)\,\rho^s_{|\bar{\beta}|,-l_0}(\bar{\beta}).
\end{align}
Taking into account the normalizations, the normalized structure constant indeed takes the form predicted by BJW conjecture (\ref{eq:C123}).

\section{Infinite volume form factors}
\label{sec:inf}
The normalized structure constant takes the same form as diagonal form factors in \textit{finite volume}. For the later case, the coefficients in front of $\rho^s$ and $\rho^c$ are identified with the diagonal form factor in \textit{infinite volume}. Keeping this analogy in mind, we also call our coefficient $F^{\mathbf{w},s}_{|\alpha|}(\alpha)$ or $F^{\mathbf{w},c}_{|\alpha|}(\alpha)$ as the infinite volume form factor. From the definition of these coefficients (\ref{eq:CF}) and (\ref{eq:CFF}), we can calculate them in terms of $p(u)$, $S(u,v)$, $h(u,v)$, $\rH^{\text{mat}}$ and their derivatives. The explicit expression becomes cumbersome very quickly.\par

For the moment, we do not have a good understanding of the structure of the infinite volume form factors. This is an interesting question to explore in the near future. One possible direction is to formulate a set of bootstrap axioms directly for the diagonal form factors and solve these axioms.\par

In the case where the light operator is BMN vacuum and the heavy operators are in the $\mathfrak{su}(2)$ sector, we can expand $F^c$ at weak coupling and compare with the known results in \cite{Hollo:2015cda} where a perfect match is found. At tree level, the infinite volume form factor $F^{c(0)}$ for $l_0=1$ is conjectured to take the following form
\begin{align}
\label{eq:FF}
F^{c(0)}_{N}(\mathbf{u})=\sigma_1^{(0)}\varphi_{12}^{(0)}\varphi_{23}^{(0)}\cdots\varphi_{N-1,N}^{(0)}+\text{permutations}
\end{align}
where $\sigma^{(0)}_i=\sigma^{(0)}(u_i)$, $\varphi^{(0)}_{ij}=\varphi^{(0)}(u_i,u_j)$ and
\begin{align}
\sigma^{(0)}(u)=\frac{1}{u^2+1/4},\qquad \varphi^{(0)}(u,v)=\frac{2}{(u-v)^2+1}
\end{align}
Interestingly, it is checked in \cite{Jiang:2015bvm} that at one loop the form (\ref{eq:FF}) still holds\footnote{We checked up to 4 excitations.} with the following corrections
\begin{align}
\sigma^{(1)}(u)=&\,\frac{1}{u^2+1/4}+\frac{8g^2 u^2}{(u^2+1/4)^3},\\\nonumber
\varphi^{(1)}(u,v)=&\,\frac{2}{(u-v)^2+1}+\frac{4g^2(u^2-v^2)}{(u^2+1/4)(v^2+1/4)((u-v)^2+1)}
\end{align}
It is possible that the structure still holds at higher loop orders\footnote{This requires to take into account also the bridge wrapping corrections.} with proper modifications of $\sigma(u)$ and $\varphi(u,v)$. This may give us some hints about the general structure of the diagonal form factors in the infinite volume and lead to more efficient ways of calculating them.

\section{Conclusions and discussions}
\label{sec:conc}
In this paper, we prove the conjecture of Bajok, Janik and Wereszczyski concerning the asymptotic volume dependence of the heavy-heavy-light structure constant at all loops in the planar $\mathcal{N}=4$ SYM theory up to mirror excitations. The proof is given for all the rank one sectors with generic non-BPS light operators.\par

In order to complete the proof, we need to take into account the bridge wrapping corrections. Once the double pole problem of the hexagon form factor approach is resolved properly, we can try to use the similar method to complete the proof. Most probably, the bridge wrapping corrections will not modify the asymptotic volume dependence but will correct the infinite volume form factors.\par

Another kind of mirror excitations give rise to physical wrapping corrections of the form $e^{-E\,L}$. For the diagonal form factor, there are conjectures of the finite volume form factor with both asymptotic volume corrections and wrapping corrections taken into account \cite{finite}. It will be very interesting to incorporate the wrapping corrections for the HHL structure constant within the hexagon approach and compare with the proposals of finite volume diagonal form factors in the literature.\par

The explicit results we have obtained so far are restricted to the $\mathfrak{su}(2)$ case and the light operator being BMN vaccum. In order to gain a general understanding of HHL structure constant, it is useful to explore other configurations. One of the most interesting case is the light operator being the dilaton. In this case, the HHL structure constant is known to be related to the derivative of the scaling dimension of the heavy operator with respect to the coupling constant $g^2$ \cite{Costa:3pt}. This allows us to make contact with the results of the spectral problem. In addition, since the relation is valid for any coupling, it may shed some light on taking into account bridge wrapping corrections.\par

It will also be interesting to perform the strong coupling expansion and compare the results with the string theory calculation in the literature \cite{Zarembo:2010ab,Costa:3pt,Bajnok:2014sza}. In this direction, one particularly interesting example is taking the giant magnon solution for the heavy operators and dilaton for the light operator.\par

Finally, the BJW conjecture only concerns the rank one sectors, namely there is only one type of excitation for the heavy operators. This is also the case that has been studied in 2d integrable field theories. A natural direction of further investigation is to study the HHL structure constant in higher rank sectors and find out the form of asymptotic volume corrections. For the operators in higher rank sectors, one needs to apply the nested Bethe ansatz and there will be richer structures to explore.

\section*{Acknowledgements}
It is my pleasure to thank Andrei Petrovskii and Laszlo Hollo for initial collaborations on the project and collaborations on related works. I'm indebted to Benjamin Basso and Shota Komatsu for many useful discussions and correspondences. I would also like to thank Zoltan Bajnok and Shota Komatsu for helpful comments on the manuscript. The work of Y.J. is partially supported by the Swiss National Science Foundation through the NCCR SwissMap.

\appendix

\section{Explicit expression for $F_2^s(u_1,u_2)$}
\label{sec:F2}
The explicit expression for the infinite volume form factor with 2 excitations is given by
\begin{align}
F_2^s(u_1,u_2)=&\,F_1^s(u_1)F_1^s(u_2)+\frac{2\mu(u_1)\mu(u_2)}{h(u_2,u_1)h(u_1,u_2)}\cos(p(u_1)-p(u_2))l_0\\\nonumber
+&\,\rH^{(0,1)}(u_1,u_1)\rH^{(0,1)}(u_2,u_2)+\rH^{(1,0)}(u_1,u_1)\rH^{(1,0)}(u_2,u_2)\\\nonumber
+&\,\frac{2\rH^{(0,1)}(u_1,u_1)\rH^{(0,1)}(u_1,u_2)}{\rH(u_1,u_2)}+\frac{2\rH^{(1,0)}(u_2,u_2)\rH^{(1,0)}(u_1,u_1)}{\rH(u_1,u_2)}\\\nonumber
+&\,\frac{2h^{(0,1)}(u_1,u_2)h^{(1,0)}(u_1,u_2)}{h(u_1,u_2)^2}-\frac{2h^{(1,1)}(u_1,u_2)}{h(u_1,u_2)}+\frac{S^{(1,1)}(u_1,u_2)}{S(u_1,u_2)}\\\nonumber
-&\,\rH^{(0,0,1,1)}(u_2,u_1,u_1,u_2)-\rH^{(1,1,0,0)}(u_1,u_2,u_2,u_1).
\end{align}

\section{Singlet state and factorization}
\label{sec:singlet}

In this appendix, we derive (\ref{eq:2gd}) in the main text. Let us first take $\chi(u)=\Phi_{1\dot{1}}(u)$ and $\bar{\chi}(v)=\Phi_{2\dot{2}}(v)$ as an example. The matrix part of the hexagon form factor $\rH_{2\gamma}(v^{2\gamma},u,\mathbf{u})$ is computed by
\begin{align}
\rH_{2\gamma}^{\chi,\text{mat}}(v^{2\gamma},u,\mathbf{u})=\langle\phi^{\dot{A}_n}\cdots\phi^{\dot{A}_1}\phi^{\dot{1}}\phi^{\dot{2}}|\mathcal{S}
|\phi^2(v^{2\gamma})\phi^1(u)\phi^{A_1}\cdots\phi^{A_n}\rangle
\end{align}
Let us focus on the part
\begin{align}
\mathcal{S}|\phi^2(v^{2\gamma})\phi^1(u)\phi^{A_1}\cdots\phi^{A_n}\rangle
\end{align}
We need to scatter all the excitations with each other. The scattering can be organized as follows, we first scatter the first two excitations in the decoupling limit $v\to u$. The result is divergent due to the kinematical pole and the residue is proportional to Beisert's singlet state \cite{Beisert:2006qh} up to a $\mathcal{Z}^-$ maker
\begin{align}
\underset{v\to u}{\text{Res}}\,\mathcal{S}_{12}|\phi^{1}_1(v^{2\gamma})\phi^2_2(u)\rangle\propto |\mathcal{Z}^-\mathbf{1}_{12}\rangle.
\end{align}
Then we scatter the singlet with rest of the excitations, which is trivial up to a scalar factor
\begin{align}
\prod_{k=1}^n \mathcal{S}_{\mathbf{1},k}|\mathcal{Z}^-\mathbf{1}_{12}\phi^{A_1}\cdots\phi^{A_n}\rangle=\prod_{k=1}^n\frac{1}{h(u^{2\gamma},u_k)h(u,u_k)}|\mathcal{Z}^-\phi^{A_1}\cdots\phi^{A_n}\mathbf{1}_{12}\rangle
\end{align}
Finally we scatter the rest of the rapidities, they contribute to $\rH_{\chi,2\gamma}^{\text{mat}}(\mathbf{u})$. It is clear that
\begin{align}
\underset{v\to u}{\text{Res}}\,\rH_{\chi,2\gamma}^{\text{mat}}(v^{2\gamma},u,\mathbf{u})\propto
\underset{v\to u}{\text{Res}}\,\rH_{\chi,2\gamma}^{\text{mat}}(v^{2\gamma},u)\cdot\frac{\rH^{\text{mat}}(\mathbf{u})}{h(u^{2\gamma},\mathbf{u})h(u,\mathbf{u})}
\end{align}
The analysis is similar for other polarizations.

The $\mathcal{Z}$ makers usually leads to some global phase factors, which needs to be taken with some care. In order to find these factors, we notice that when forming the singlet state, there is a difference of $\mathcal{Z}^-$ maker between bosonic and fermionic excitations
\begin{align}
\mathcal{S}_{12}|\phi^a_1\phi^b_2\rangle\sim|\mathcal{Z}^-\mathbf{1}_{21}\rangle,\qquad \mathcal{S}_{12}|\psi^\alpha_1\psi^\beta_2\rangle\sim|\mathbf{1}_{12}\rangle
\end{align}
also there is a $\mathcal{Z}^+$ marker difference between the bosonic and fermionic excitations in the singlet state
\begin{align}
\label{eq:fft}
|\mathbf{1}_{12}\rangle=\frac{\alpha}{\gamma_1\gamma_2}\left(\frac{x_1^+}{x_1^-}-1 \right)\epsilon_{ab}|\mathcal{Z}^+\phi^a_1\phi^b_2\rangle+\epsilon_{\alpha\beta}|\psi_1^\alpha\psi_2^\beta\rangle.
\end{align}
For $\chi,\bar{\chi}$ being \textbf{scalars}, one scatters the first two bosonic excitations $\phi^a$ and form a singlet with $\mathcal{Z}^-$ marker, then move the singlet to the rightmost, finally contract the singlet with the scalar excitations of the right sector where we need to take into account the $\mathcal{Z}^+$ marker. We need to move the $\mathcal{Z}^+$ marker to the leftmost in order to pull it out. The $\mathcal{Z}^+$ and $\mathcal{Z}^-$ markers cancel each other. However, when moving the $\mathcal{Z}^+$ markers to the leftmost, we pick up the phase factor $e^{iP}$ by the rule of moving makers.\par

For $\chi,\bar{\chi}$ being \textbf{derivatives}, one scatters the fermionic excitations $\psi^\alpha$ and form a singlet. Then move the singlet to the rightmost and finally contract the singlet with fermionic excitations of the right sector. No markers are involved in the process, hence the phase factor is $1$.\par

For $\chi,\bar{\chi}$ being \textbf{fermions}, there are two types of process. The first corresponds to scattering the scalar excitations and contract the singlet with fermionic excitations in the right sector. This involves a $\mathcal{Z}^-$ maker on the left and no $\mathcal{Z}^+$ maker. Pulling it out, we get a phase factor $e^{\frac{i}{2}P}$. The second case corresponds to scattering the fermionic excitations and contract the singlet with bosonic excitations. This involves a $\mathcal{Z}^+$ marker on the rightmost and no $\mathcal{Z}^-$ markers. Moving the $\mathcal{Z}^+$ marker to the leftmost picks up a phase $e^{iP}$, pulling it out gives another phase $e^{-\frac{i}{2}P}$. In total the phase factor is $e^{\frac{i}{2}P}$. To summarize, the phase factors for the three kind of excitations are given by
\begin{align}
\label{eq:phase}
\Phi:\quad e^{iP},\qquad \mathcal{D}:\quad 1,\qquad \Psi:\quad e^{\frac{i}{2}P}.
\end{align}
Combining (\ref{eq:fft}) and (\ref{eq:phase}), we obtain (\ref{eq:2gd}) in the main text.

\section{The ratio of phase factors}
\label{sec:ratio}
In this appendix, we calculate the ratio of phase factors in (\ref{eq:ratio}) of the main text.
\subsection{Scalars}
For scalar excitations, we can take $\chi=\Phi_{1\dot{1}}$ and $\bar{\chi}={\Phi}_{2\dot{2}}$ and consider the following hexagon form factor
\begin{align}
\rH_{\Phi}=\langle\mathfrak{h}|\Phi_{1\dot{1}}(u)\mathcal{X}_{\mathbf{A}\dot{\mathbf{A}}}(\mathbf{u})\rangle|{\Phi}_{2\dot{2}}(v)\rangle|0\rangle
\end{align}
The phase factor contains three parts
\begin{itemize}
\item Phases come from changing from spin chain frame to string frame before crossing;
\item Phases come from crossing transformation of ${\Phi}_{2\dot{2}}(v)$;
\item Phases come from changing from string frame to spin chain frame after crossing.
\end{itemize}
The first part is the same for both $2\gamma$ and $-4\gamma$ transformations. The second part is $-1$ for $2\gamma$ transformation and $1$ for $-4\gamma$ transformation. In order to find the ratio of the two phase factors, it is enough to consider only the third part. Let us remind here that the transformation rules between spin chain frame and string frame for the derivatives, scalars and fermions are given by\footnote{We thank S. Komatsu for informing us the transformation rule for the fermions.} \cite{Basso:2015zoa}
\begin{align}
\mathcal{D}_{\text{string}}=\mathcal{D}_{\text{spin}},\quad\Phi_{\text{string}}=\sqrt{Z}\Phi_{\text{spin}}\sqrt{Z},\quad
\Psi_{\text{string}}=Z^{1/4}\Psi_{\text{spin}}Z^{1/4}.
\end{align}
\paragraph{The $2\gamma$ transformation}
\begin{align}
\label{eq:FA}
\langle\mathfrak{h}|\Phi_{2\dot{2}}(v^{2\gamma})\Phi_{1\dot{1}}(u)\mathcal{X}_{\mathbf{A}\dot{\mathbf{A}}}(\mathbf{u})\rangle|0\rangle|0\rangle_{\text{string}}
=F_A\,\langle\mathfrak{h}|\sqrt{Z}\Phi_{2\dot{2}}(v^{2\gamma})Z\Phi_{1\dot{1}}(u)\sqrt{Z}Z^n\mathcal{X}_{\mathbf{A}\dot{\mathbf{A}}}(\mathbf{u})\rangle|0\rangle|0\rangle_{\text{spin}}
\end{align}
Here $F_A$ is the phase factor coming from moving all the $Z$-markers of $\mathcal{X}_{\mathbf{A}\dot{\mathbf{A}}}(\mathbf{u})$ to the left. Since we allow any kind of excitations, $n$ does not have to be equal to $N$ and not even have to be an integer.
We can then move all the $Z$-markers to the leftmost and then pull them out using the rule
\begin{align}
\langle\mathfrak{h}|Z^n\psi\rangle=z^n\langle\mathfrak{h}|\psi\rangle,\quad z=e^{-ip/2}
\end{align}
where $p$ is the \emph{total momentum} of the state $|\psi\rangle$. The result is given by
\begin{align}
\label{eq:2g}
\langle\mathfrak{h}|\Phi_{2\dot{2}}(v^{2\gamma})\Phi_{1\dot{1}}(u)\mathcal{X}_{\mathbf{A}\dot{\mathbf{A}}}(\mathbf{u})\rangle|0\rangle|0\rangle_{\text{string}}
=&\,\langle\mathfrak{h}|\Phi_{2\dot{2}}(v^{2\gamma})\Phi_{1\dot{1}}(u)\mathcal{X}_{\mathbf{A}\dot{\mathbf{A}}}(\mathbf{u})\rangle|0\rangle|0\rangle_{\text{spin}}\\\nonumber
&\times e^{\frac{i}{2}(n-1)p_1-\frac{i}{2}(n+1)p_2-\frac{i}{2}(n+2)P}\,F_A.
\end{align}

\paragraph{The $-4\gamma$ transformation.}
\begin{align}
\langle\mathfrak{h}|\Phi_{1\dot{1}}(u)\mathcal{X}_{\mathbf{A}\dot{\mathbf{A}}}(\mathbf{u})\Phi_{2\dot{2}}(v^{-4\gamma})\rangle|0\rangle|0\rangle_{\text{string}}=
F_A\,\langle\mathfrak{h}|\sqrt{Z}\Phi_{1\dot{1}}(u)\sqrt{Z} Z^n\mathcal{X}_{\mathbf{A}\dot{\mathbf{A}}}(\mathbf{u})\sqrt{Z}\Phi_{2\dot{2}}(v^{-4\gamma})\sqrt{Z}\rangle|0\rangle|0\rangle_{\text{spin}}.
\end{align}
where $F_A$ is the same phase factor as in (\ref{eq:FA}). By moving and pulling out the $Z$-markers, we obtain
\begin{align}
\label{eq:4g}
\langle\mathfrak{h}|\Phi_{1\dot{1}}(u)\mathcal{X}_{\mathbf{A}\dot{\mathbf{A}}}(\mathbf{u})\Phi_{2\dot{2}}(v^{-4\gamma})\rangle|0\rangle|0\rangle_{\text{string}}
=&\,\langle\mathfrak{h}|\Phi_{1\dot{1}}(u)\mathcal{X}_{\mathbf{A}\dot{\mathbf{A}}}(\mathbf{u})\Phi_{2\dot{2}}(v^{-4\gamma})\rangle|0\rangle|0\rangle_{\text{spin}}\\\nonumber
&\,\times e^{\frac{i}{2}(n+1)p_1-\frac{i}{2}(n+1)p_2-\frac{i}{2}nP} F_A.
\end{align}
From (\ref{eq:2g}) and (\ref{eq:4g}) and taking into account the relative minus sign from crossing transformation, it is clear that
\begin{align}
\frac{\texttt{phase}_{2\gamma}}{\texttt{phase}_{-4\gamma}}=-e^{-ip_1-iP}.
\end{align}

\subsection{Derivatives}
For derivatives, we take $\chi=\mathcal{D}_{3\dot{3}}$, $\bar{\chi}=\mathcal{D}_{4\dot{4}}$ and consider the following configuration
\begin{align}
\rH_D=\langle\mathfrak{h}|\mathcal{D}_{3\dot{3}}(u)\mathcal{X}_{\mathbf{A}\dot{\mathbf{A}}}(\mathbf{u})\rangle|{\mathcal{D}}_{4\dot{4}}(v)\rangle|0\rangle.
\end{align}
Again we only consider the third step of changing back from string frame to spin chain frame since we are considering the ratios.
\paragraph{$2\gamma$ transformation}
\begin{align}
\label{eq:2gD}
\langle\mathfrak{h}|{\mathcal{D}}_{4\dot{4}}(v^{2\gamma})\mathcal{D}_{3\dot{3}}(u)\mathcal{X}_{\mathbf{A}\dot{\mathbf{A}}}(\mathbf{u})\rangle|0\rangle|0\rangle_{\text{string}}
=&\,\langle\mathfrak{h}|{\mathcal{D}}_{4\dot{4}}(v^{2\gamma})\mathcal{D}_{3\dot{3}}(u)\mathcal{X}_{\mathbf{A}\dot{\mathbf{A}}}(\mathbf{u})\rangle|0\rangle|0\rangle_{\text{spin}}\\\nonumber
\times&\,e^{i(p_1-p_2)n-\frac{i}{2}(p_1-p_2+P)n}\,F_A
\end{align}

\paragraph{$-4\gamma$ transformation}
\begin{align}
\label{eq:4gD}
\langle\mathfrak{h}|\mathcal{D}_{3\dot{3}}(u)\mathcal{X}_{\mathbf{A}\dot{\mathbf{A}}}(\mathbf{u}){\mathcal{D}}_{4\dot{4}}(v^{-4\gamma})\rangle|0\rangle|0\rangle_{\text{string}}
=&\,\langle\mathfrak{h}|\mathcal{D}_{3\dot{3}}(u)\mathcal{X}_{\mathbf{A}\dot{\mathbf{A}}}(\mathbf{u}){\mathcal{D}}_{4\dot{4}}(v^{-4\gamma})\rangle|0\rangle|0\rangle_{\text{spin}}\\\nonumber
\times&\,e^{ip_1 n-\frac{i}{2}(p_1+p_2+P)n}\,F_A.
\end{align}

Comparing (\ref{eq:2gD}) and (\ref{eq:4gD}), we conclude that for the derivatives
\begin{align}
\frac{\texttt{phase}_{2\gamma}^D}{\texttt{phase}_{-4\gamma}^D}=-1.
\end{align}

\subsection{Fermions}
For fermions, we take $\chi=\Psi_{1\dot{3}}$ and $\bar{\chi}={\Psi}_{4\dot{2}}$ and consider the following configuration
\begin{align}
\rH_\Psi=\langle\mathfrak{h}|\Psi_{1\dot{3}}(u)\mathcal{X}_{\mathbf{A}\dot{\mathbf{A}}}(\mathbf{u})\rangle|{\Psi}_{4\dot{2}}(v)\rangle|0\rangle.
\end{align}
\paragraph{$2\gamma$ transformation}
\begin{align}
\langle\mathfrak{h}|\Psi_{2\dot{4}}(v^{2\gamma})\Psi_{1\dot{3}}(u)\mathcal{X}_{\mathbf{A}\dot{\mathbf{A}}}(\mathbf{u})\rangle|0\rangle|0\rangle_{\text{string}}
=&\,\langle\mathfrak{h}|\Psi_{2\dot{4}}(v^{2\gamma})\Psi_{1\dot{3}}(u)\mathcal{X}_{\mathbf{A}\dot{\mathbf{A}}}(\mathbf{u})\rangle|0\rangle|0\rangle_{\text{spin}}\\\nonumber
\times&\,e^{\frac{i}{4}p_1-\frac{3i}{4}p_2+i(p_1-p_2)n-\frac{i}{2}(p_1-p_2+P)(n+1)}\,F_A.
\end{align}
\paragraph{$-4\gamma$ transformation}
\begin{align}
\langle\mathfrak{h}|\Psi_{1\dot{3}}(u)\mathcal{X}_{\mathbf{A}\dot{\mathbf{A}}}(\mathbf{u}){\Psi}_{4\dot{2}}(v^{2\gamma})\rangle|0\rangle|0\rangle_{\text{string}}
=&\,\langle\mathfrak{h}|\Psi_{1\dot{3}}(u)\mathcal{X}_{\mathbf{A}\dot{\mathbf{A}}}(\mathbf{u}){\Psi}_{4\dot{2}}(v^{2\gamma})\rangle|0\rangle|0\rangle_{\text{spin}}\\\nonumber
&\times e^{\frac{i}{4}(p_1+p_2)+ip_1 n+\frac{i}{2}(p_1+P)-\frac{i}{2}(p_1+p_2+P)(n+1)}\,F_A.
\end{align}
Comparing the two results, we obtain
\begin{align}
\frac{\texttt{phase}_{2\gamma}^\Psi}{\texttt{phase}_{-4\gamma}^\Psi}=-e^{-\frac{i}{2}(p_1+P)}
\end{align}

\section{Computation of $n_\chi$}
\label{sec:nchi}
In this appendix, we compute $n_\chi$ for different polarizations.
\begin{itemize}
\item $\chi=\Phi_{1\dot{1}}$, $\bar{\chi}=\Phi_{2\dot{2}}$
\begin{align}
\rH_{\chi}^{\text{mat}}(v^{2\gamma},u)=&\,-\frac{1}{2}(A(v^{2\gamma},u)+B(v^{2\gamma},u))\\\nonumber
i\underset{v\to u}{\text{Res}}\,\rH_\chi^{\text{mat}}(v^{2\gamma},u)=&\,-\frac{i}{2}\underset{v\to u}{\text{Res}}\,B(v^{2\gamma},u)=e^{ip}\mu(u).
\end{align}
Therefore $n_\chi=1$.
\item $\chi=\Phi_{1\dot{2}}$, $\bar{\chi}=\Phi_{2\dot{1}}$
\begin{align}
\rH_{\chi}^{\text{mat}}(v^{2\gamma},u)=&\,-\frac{1}{2}(A(v^{2\gamma},u)-B(v^{2\gamma},u))\\\nonumber
i\underset{v\to u}{\text{Res}}\,\rH_\chi^{\text{mat}}(v^{2\gamma},u)=&\,\frac{i}{2}\underset{v\to u}{\text{Res}}\,B(v^{2\gamma},u)=-e^{ip}\mu(u).
\end{align}
Therefore $n_\chi=-1$.
\item $\chi=\Phi_{3\dot{3}}$, $\bar{\chi}=\Phi_{4\dot{4}}$
\begin{align}
\rH_{\chi}^{\text{mat}}(v^{2\gamma},u)=&-\frac{1}{2}(D(v^{2\gamma},u)+E(v^{2\gamma},u))\\\nonumber
i\underset{v\to u}{\text{Res}}\,\rH_\chi^{\text{mat}}(v^{2\gamma},u)=&\,-\frac{i}{2}\underset{v\to u}{\text{Res}}\,E(v^{2\gamma},u)=\mu(u)
\end{align}
therefore $n_\chi=1$
\item $\chi=\Phi_{3\dot{4}}$, $\bar{\chi}=\Phi_{4\dot{3}}$
\begin{align}
\rH_{\chi}^{\text{mat}}(v^{2\gamma},u)=&-\frac{1}{2}(D(v^{2\gamma},u)-E(v^{2\gamma},u))\\\nonumber
i\underset{v\to u}{\text{Res}}\,\rH_\chi^{\text{mat}}(v^{2\gamma},u)=&\,\frac{i}{2}\underset{v\to u}{\text{Res}}\,E(v^{2\gamma},u)=-\mu(u)
\end{align}
therefore $n_\chi=-1$.
\item $\chi=\Psi_{1\dot{3}}$, $\bar{\chi}={\Psi}_{2\dot{4}}$,
\begin{align}
\rH_{\chi}^{\text{mat}}(v^{2\gamma},u)=&\,-\frac{1}{2}C(v^{2\gamma},u)e^{\frac{i}{2}(p(u)-p(v))}\\\nonumber
i\underset{v\to u}{\text{Res}}\,\rH_\chi^{\text{mat}}(v^{2\gamma},u)=&\,\frac{i}{2}\underset{v\to u}{\text{Res}}\,C(v^{2\gamma},u)=\mu(u)\,e^{\frac{i}{2}p}
\end{align}
therefore $n_\chi=1$. The rest three fermionic excitations also gives $n_{\chi}=\pm1$.
\end{itemize}
The result is summarized in the following table
\begin{center}
\begin{tabular}{|c|c|c|c|c|c|c|c|c|}
  \hline
  $\chi$       & $\Phi_{1\dot{1}}$            & $\Phi_{1\dot{2}}$            & $\mathcal{D}_{3\dot{3}}$            &$\mathcal{D}_{3\dot{4}}$  & $\Psi_{1\dot{3}}$ & $\Psi_{2\dot{3}}$ & $\Psi_{1\dot{4}}$ & $\Psi_{2\dot{4}}$ \\
  \hline
  $\bar{\chi}$ & ${\Phi}_{2\dot{2}}$ & ${\Phi}_{2\dot{1}}$ & ${\mathcal{D}}_{4\dot{4}}$ &${\mathcal{D}}_{4\dot{3}}$ & ${\Psi}_{4\dot{2}}$ & ${\Psi}_{4\dot{1}}$ & ${\Psi}_{3\dot{2}}$ & ${\Psi}_{3\dot{1}}$ \\
  \hline
  $n_{\chi}$   & $1 $                         & $-1$                         & $1$                                 & $-1$ & $1$ & $-1$ & $-1$ & $1$ \\
  \hline
\end{tabular}
\end{center}

\end{document}